\documentclass{aa}

\newcommand{\kepler}{{\it Kepler}}

\usepackage{hyperref}
\usepackage{graphicx}
\usepackage{txfonts}
\usepackage{breakurl}

\graphicspath{ {figures/} }

\begin{document}

\title{Recurrent mini-outbursts and a magnetic
 white dwarf in the symbiotic system FN\,Sgr}
\titlerunning{FN\,Sgr with magnetic WD}
\authorrunning{Magdolen et al.}

\author{J.~Magdolen \inst {1}, A.~Dobrotka \inst {1}, M.~Orio \inst {2,3}, J. Miko{\l}ajewska \inst {4}, A.~Vanderburg \inst {6}, B.~Monard \inst {5}, R. Aloisi \inst {2} and P.~Bez\'ak \inst {1}}

\institute{Advanced Technologies Research Institute, Faculty of Materials Science and Technology in Trnava, Slovak University of Technology in Bratislava, Bottova 25, 917 24 Trnava, Slovakia
\and
Department of Astronomy, University of Wisconsin 475 N. Charter Str. Madison, WI 53706
\and
INAF - Astronomical Observatory Padova, vicolo dell'Osservatorio 5, 35122 Padova, Italy
\and
Nicolaus Copernicus Astronomical Center, Polish Academy of Sciences, Bartycka 18, 00-716 Warsaw, Poland
\and
Kleinkaroo Observatory, Calitzdorp, Western Cape, South Africa
\and
Department of Physics and Kavli Institute for Astrophysics and Space Research, Massachusetts Institute of Technology, 77 Massachusetts Avenue, Cambridge, MA 02139, USA
}

\date{Received / Accepted}

\abstract
{}
{We investigated the optical variability of the symbiotic binary FN\,Sgr, with photometric monitoring during $\simeq$55 years and with a high-cadence \kepler\ light curve lasting 81 days.}
{The data obtained in the V and I bands were reduced with standard photometric methods. The \kepler\ data were divided into subsamples and analyses with the Lomb-Scargle algorithm.}
{The V and I band light curves showed a phenomenon never before observed with such recurrence in any symbiotic system, namely short outbursts, starting between orbital phase 0.3 and 0.5 and lasting about a month, with a fast rise and a slower decline, and amplitude of 0.5-1 mag. In the \kepler\ light curve we discovered three frequencies with sidebands. We attribute a stable frequency of 127.5 d$^{-1}$ (corresponding to an 11.3 minutes period) to the white dwarf rotation. We suggest that this detection probably implies that the white dwarf accretes through a magnetic stream, like in intermediate polars. The small outbursts may be ascribed to the stream-disc interaction. Another possibility is that they are due to localized thermonuclear burning, perhaps confined by the magnetic field, like recently inferred in intermediate polars, albeit on different timescales. We measured also a second frequency around 116.9 d$^{-1}$ (corresponding to about 137 minutes), which is much less stable and has a drift. It may be due to rocky detritus around the white dwarf, but it is more likely to be caused by an inhomogeneity in the accretion disk. Finally, there is a third frequency close to the first one that appears to correspond to the beating between the rotation and the second frequency.}
{}

\maketitle

\section{Introduction}
\label{introduction}

Symbiotic stars are binary systems, usually comprised of a white dwarf (WD), main sequence, or a neutron star and an evolved companion, namely a red giant (S-type systems), or an asymptotic giant branch star or a Mira variable (D-type systems). The matter from the companion is accreted by the compact object via a stellar wind or Roche lobe overflow, in many cases creating an accretion disk. The orbital periods are $\sim$ 200-1000 days for S-type systems and $\gtrsim 50$ years for D-type systems (\citealt{mikolajewska2012}).

FN\,Sgr has been mainly studied spectroscopically \citep{barba1992,munari1994,Brandi2005}. \citet{munari1994} found some evidence that the WD may be steadily burning accreted hydrogen near its surface. \citet{Brandi2005} presented long-term photometric data over 30 years and derived the orbital period, 568.3 days, supported also by the spectroscopic analysis. As we discuss below, we were able to slightly revise this period in this paper. \citet{Brandi2005} determined the interstellar extinction (E(B-V)=0.2$\pm$0.1), a distance of about 7$^{+1}_{-2}$ kpc assuming that the giant fills or almost fills the Roche lobe, and orbital inclination of 80$^{\rm o}$. These authors found that FN Sgr is an S-type symbiotic, composed of an M 5-type giant of 1.5\,M$_\odot$ and radius 140\,R$_\odot$ and a hot WD of 0.7\,M$_\odot$, with binary separation 1.6\,AU. They also concluded that the mass transfer process is most likely caused by the Roche lobe overflow of the red giant. An accretion disk is likely to be present and it is most likely the source of the spectral features. \citet{Brandi2005} discuss the double-temperature structure of the hot component, as already observed in other symbiotics, possibly due to a geometrically and optically thick accretion disk. \citet{Brandi2005} discovered a 1996 optical outburst with a 2.5 mag amplitude that lasted until early 2001. Even if outbursts are not unusual in symbiotics, the temperature and luminosity evolution of the hot component during this outburst, were not consistent with an accretion disk instability and were difficult to explain \citep{Brandi2005}.

Since the only high-time resolution light curve lasting only 2.8 hours (\citealt{sokoloski1999}) is not suitable for detailed timing analysis, we propose to obtain a high cadence \kepler\ light curve, which was measured over 81 days, between 2015 October and December, in \kepler\ K2 field 7.

\section{Observations}

\begin{figure}
\begin{center}
\includegraphics[width=90mm]{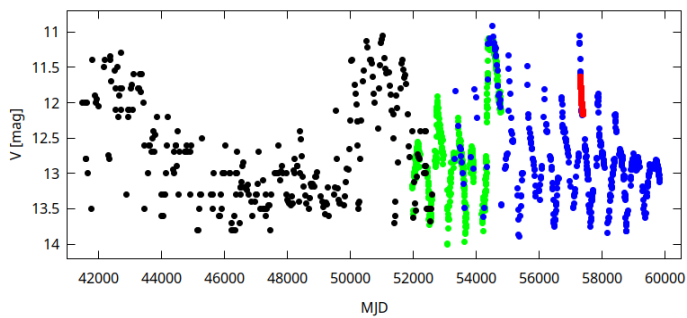}
\caption{Optical light curve of FN Sgr: in black data from \citep{Brandi2005} in green the ASAS data, and in blue our new data. The \kepler\ light curve is also shown in red, offset vertically by 12 magnitudes for comparison with the V-band data.}
\label{optical_plus_kepler}
\end{center}
\end{figure}

The optical data in Fig.~\ref{optical_plus_kepler} includes data from \citet{Brandi2005}, from the All-Sky-Automated Survey \citep[ASAS, see][]{gromadzki2013} and new photometry we obtained with the 35cm Meade RCX400 telescope at the Kleinkaroo Observatory using a SBIG ST8-XME CCD camera and V and Ic filters. The new V light curve we obtained starts at MJD\footnote{MJD = JD - 2400000} 53308, but the Ic data cover a shorter period, starting at MJD 56233. Each observation was the result of several individual exposures, calibrated (dark-subtraction and flat-fielding) and stacked. The magnitudes were derived from differential photometry, with nearby reference stars, using the single image mode of the AIP4 image processing software. The photometric accuracy of the derived magnitudes is better than 0.1 mag.

\begin{figure}
\begin{center}
\includegraphics[width=90mm]{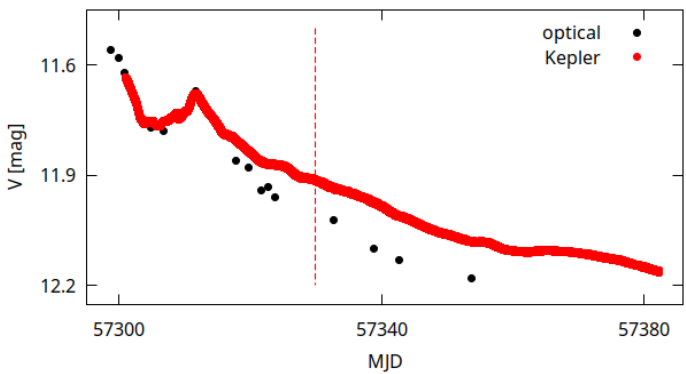}
\caption{\kepler\ light curve (red points) with optical data from fig.~\ref{optical_plus_kepler}. The red dashed line shows the superior conjunction of the red giant.}
\label{optical_plus_kepler_2}
\end{center}
\end{figure}

The \kepler\ light curve was measured on 2015-10-5 (EPIC 218331937). It is shown in Fig.~\ref{optical_plus_kepler_2} in red. Before flux normalization, systematic corrections were applied following \citet{vanderburg2014} and \citet{vanderburg2016} \footnote{The \kepler\ team evaluates the contribution of scattered background light and subtracts it at the pixel level. The background is a large source of photon noise. For faint stars, where the background is much larger than the star flux, the estimated brightness may be negative} namely background and barycentric correction. We transformed the flux to magnitudes using the equation: $m = -2.5\log(f/f_0)$, where $f$ corresponds to the flux value and $f_0$ was set to 1 (not standardized \kepler's response function). A vertical offset of 12 was used to compare the \kepler \ and V magnitudes.
\begin{figure}
\begin{center}
\includegraphics[width=90mm]{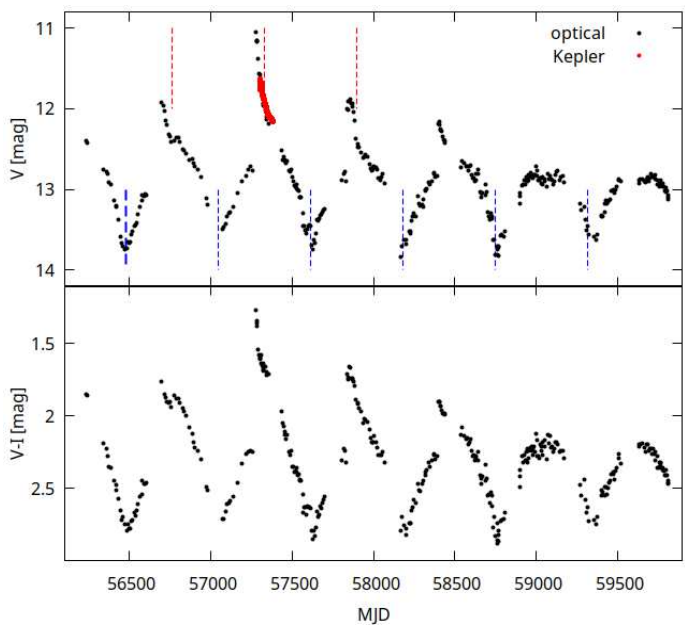}
\caption{The upper panel shows a selected time interval from Fig.~\ref{optical_plus_kepler}, and the bottom panel shows the V-I index. The blue dashed lines correspond to the inferior conjunction, the red ones to the superior conjunction of the red giant. The blue thick dashed line is the reference optical minimum determined by polynomial fitting.}
\label{optical_plus_kepler_3}
\end{center}
\end{figure}

\section{Long-term light curve}

The optical V light curve, shown in Fig.~\ref{optical_plus_kepler}, with all the data we could gather at this stage, covers almost 55 years. The minima represent the inferior spectroscopic conjunction of the red giant \citep{Brandi2005}. We used all available photometric measurements, including the additional data shown in Fig. 2 \citep{gromadzki2013} and revised the ephemeris of the minimum as follows:
\begin{equation*}
{\rm JD}({\rm MIN}) = 2450260+(567.3 \pm 0.3) \times E,
\end{equation*}

where $E$ is the number of orbital cycles, implying that the orbital period is one day shorter than calculated earlier, 567.3 days. The blue dashed lines in Fig.~\ref{optical_plus_kepler_3} show the minima with a reference minimum determined using a 6th order polynomial (indicated as a blue thick line). The corresponding superior conjunctions are indicated by red dashed lines.

In Fig.~\ref{optical_plus_kepler} and in the selected interval in Fig.~\ref{optical_plus_kepler_3} it is clear that small amplitude flares, of amplitude 0.5-1\,mag, which we will call here mini-outbursts to differentiate them from major ones like the 1997-1998 event, occurred between 2001 and 2019 and seem to have ceased in 2020-2021. The ASAS light curve \citep{gromadzki2013} indicates that the mini-outbursts occurred since 2001.

Fig.~\ref{optical_plus_kepler_3} shows that these outbursts have a sharp rise ($\sim 10$ days) and slower decline. They also seem to be orbitally phase-locked. The rise usually starts near phase 0.3 and never after phase 0.5. The bottom panel of Fig.~\ref{optical_plus_kepler_3} shows that the peak wavelength during the mini-outbursts was shifted towards higher energies since V-I clearly decreases in magnitude in the flares.

Since 2019 the mini-outburst activity has ceased, and the light curves in addition to the deep eclipses show secondary minima, which are most likely due to ellipsoidal variability. This is particularly evident in the I light curve, in which the red giant significantly contributes to the continuum and confirms the conclusion by \citet{Brandi2005} that the red giant fills, or almost fills, its Roche lobe, as these authors inferred from the analysis of the shape and duration of the well-defined eclipses during the large outburst in 1996-2001. If the Roche lobe is filled, a persistent accretion disk should be present in this system.

\section{Timing analysis of the Kepler light curve}

Our \kepler\ light curve was observed over 81 days with a cadence of almost 1 minute. Such a high-quality and long light curve is ideal for period analysis. The 81 days Kepler run, shown in detail in Fig.~\ref{optical_plus_kepler_2}, occurred during the decay from a mini-outburst, as shown in Fig.~\ref{optical_plus_kepler} and \ref{optical_plus_kepler_3}. Thus, in addition to orbital variability a decay after the flare is observed, with a few outlier points due to cosmic rays. To eliminate these effects, first the Hampel filter\footnote{Python hampel library \url{https://github.com/MichaelisTrofficus/hampel_filter}.} was used to detect and remove outliers. By specifying the size of the window as 61 (30 points on each side + 1 central point), the central point inside this window was replaced by the window's median value, if it differed more than $5\sigma$ from the window's median. Next, in order to detrend the light curve, we used a moving window median\footnote{SciPy Python's package \url{https://scipy.org/}.} with window size 201 (100 points each side + 1 central point) because a polynomial detrend performed poorly for such a long observation. The window slides over the entire light curve, point by point, where the central point is replaced with a median value calculated over the window.

% Next, in order to detrend the light curve we used a moving window median\footnote{SciPy
% Python's package \url{https://scipy.org/}.}, because a polynomial detrend performed poorly
% for such a long observation. The derived median values were calculated over a sliding window % of size 201 (100 points each side + 1 central point) across the light curve.

Subsequently, we applied the Lomb-Scargle (LS) method by \cite{scargle1982} to the processed light curve\footnote{Astropy Python's package \url{https://docs.astropy.org/en/stable/index.html} was calculated with normalization set to "standard".}.

\begin{figure}
\begin{center}
\includegraphics[width=90mm]{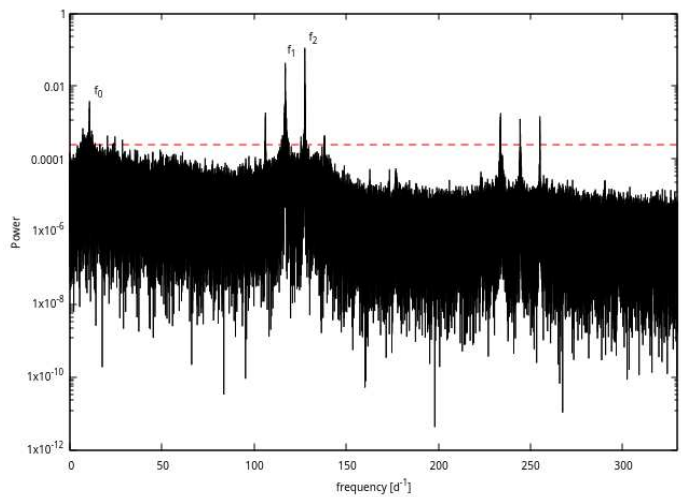}
\caption{LS periodogram with Power in logarithmic scale. The red dashed line indicates the 90\% confidence level.}
\label{LS_periodogram}
\end{center}
\end{figure}

\begin{figure}
\begin{center}
\includegraphics[width=90mm]{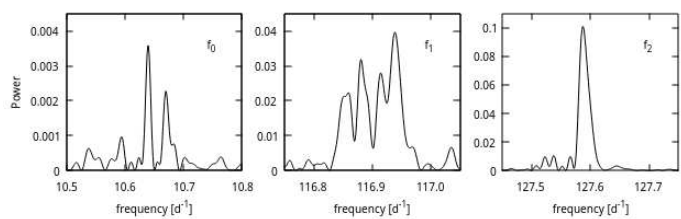}
\caption{The peak pattern of the $f_0$, $f_1$ and $f_2$ frequencies given by the LS periodogram.}
\label{main_frequencies}
\end{center}
\end{figure}

The resulting periodogram shows several peaks above the 90-\% confidence level (Fig.~\ref{LS_periodogram}). We show in Figure \ref{main_frequencies} that zooming at the most significant frequencies, namely 10.5 d$^{-1}$ ($f_0$), $116.9$ d$^{-1}$ ($f_1$) and $127.5$ d$^{-1}$ ($f_2$) we observe a multi-peak pattern for $f_0$ and $f_1$. Such a drift in frequency is typical for a quasi-periodic signal, therefore these frequencies may not be stable. On the contrary, $f_2$ exhibits only a single dominant peak, suggesting a stable frequency.

\begin{figure}
\begin{center}
\includegraphics[width=90mm]{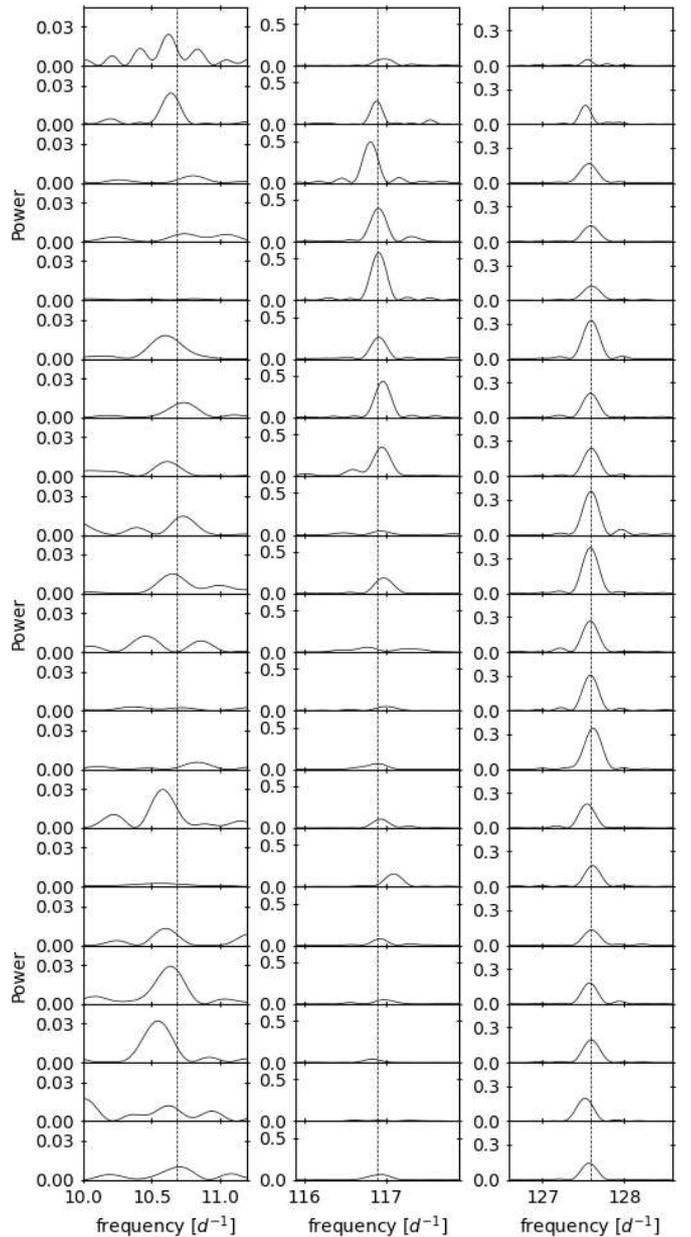}
\caption{LS periodogram per portion, obtained by dividing the Kepler light curve into 20 equally spaced sub-samples. The first column represents $f_0$, the second $f_1$, and the third $f_2$. The vertical lines indicate the mean value of $f_1$ and $f_2$, or $\Delta$ when $f_0$ was computed as the difference between the two frequencies.}
\label{frequencies_zoom}
\end{center}
\end{figure}

To determine the origin and the exact behaviour of these frequencies, we split the corrected light curve into 20 portions; 2 equally large parts between days 0 and 10, and 18 equally large parts observed between days 10 and 80. This selection was made in order to distinguish the U-shaped trend at the beginning of the light curve. Subsequently, LS periodograms were created for each portion. First, we inspected the $f_0$, $f_1$, and $f_2$ frequencies. By fitting a Gaussian (or multiple Gaussians where needed), we estimated the frequencies and their uncertainties, given in Table \ref{table_frequencies}. The missing values in the $f_0$ column indicate that the peak confidence fell below 90\%. Two frequency values instead are listed when more peaks were present, with two close frequencies. Then, we calculated the mean (with the whole light curve) of the two main frequencies $f_1$ and $f_2$. By subtracting them, we found the difference $\Delta(f) \approx 10.66$ d$^{-1}$, which is very close to the $f_0$ peak. The LS periodogram for each portion of the light curve, with mean frequencies indicated by vertical lines, is shown for $f_1$ and $f_2$ in Figure \ref{frequencies_zoom}, and the $\Delta$ value for $f_0$ is also indicated.

By examining Table \ref{table_frequencies} and Figure $\ref{frequencies_zoom}$, we observe the trends of the frequencies. The lower frequency $f_0$ exhibits a visible drift around the $\Delta$ value. Moreover, the frequency disappeared at certain times, as its power fell below the 90\% confidence level. A similar behaviour is observed for $f_1$, where the peaks drift around the mean value. On the other side, we can see an obvious decline in power in the second half of the light curve, although it is still above the selected confidence level. Only the $f_2$ frequency appears to be stable, its drift around the mean is minimal, and so is the decrease in power. In Sect. \ref{sec_discussion} we discuss the interpretation of the results in more detail.
\begin{table}
\caption{Gaussian frequencies fit. The empty $f_0$ values indicate the absence of any peak over the 90\% confidence level. The 4th, 11th and 19th snapshot manifest peak splitting at $f_0$, with two close frequencies above 90\% confidence level.}
\begin{center}
\begin{tabular}{rccc}
\hline
\hline
s & $f_0$ & $f_1$\  & $f_2$\\
  & $(d^{-1})$ & $(d^{-1})$ & $(d^{-1})$\\
\hline
1 & $10.63 \pm 0.06$ & $116.97 \pm 0.10$ & $127.55 \pm 0.05$\\
2 & $10.64 \pm 0.06$ & $116.88 \pm 0.07$ & $127.53 \pm 0.06$\\
3 & $10.81 \pm 0.09$ & $116.81 \pm 0.09$ & $127.56 \pm 0.10$\\
4 & $10.74 \pm 0.10$ & $116.90 \pm 0.09$ & $127.59 \pm 0.09$\\
  & $11.05 \pm 0.09$ & &\\
5 & - & $116.91 \pm 0.09$ & $127.60 \pm 0.10$\\
6 & $10.61 \pm 0.12$ & $116.91 \pm 0.09$ & $127.59 \pm 0.09$\\
7 & $10.73 \pm 0.10$ & $116.96 \pm 0.09$ & $127.59 \pm 0.09$\\
8 & $10.62 \pm 0.09$ & $116.95 \pm 0.11$ & $127.60 \pm 0.09$\\
9 & $10.74 \pm 0.09$ & $116.94 \pm 0.10$ & $127.59 \pm 0.09$\\
10 & $10.65 \pm 0.10$ & $116.97 \pm 0.11$ & $127.59 \pm 0.09$\\
11 & $10.46 \pm 0.08$ & $116.77 \pm 0.10$ & $127.59 \pm 0.09$\\
   & $10.87 \pm 0.08$ & &\\
12 & - & $117.01 \pm 0.12$ & $127.59 \pm 0.06$\\
13 & $10.84 \pm 0.09$ & $116.87 \pm 0.14$ & $127.61 \pm 0.10$\\
14 & $10.59 \pm 0.09$ & $116.93 \pm 0.10$ & $127.55 \pm 0.09$\\
15 & - & $117.09 \pm 0.10$ & $127.61 \pm 0.09$\\
16 & $10.60 \pm 0.08$ & $116.92 \pm 0.09$ & $127.60 \pm 0.09$\\
17 & $10.63 \pm 0.10$ & $116.99 \pm 0.11$ & $127.58 \pm 0.09$\\
18 & $10.54 \pm 0.10$ & $116.83 \pm 0.10$ & $127.59 \pm 0.09$\\
19 & $10.62 \pm 0.09$ & $116.62 \pm 0.09$ & $127.52 \pm 0.09$\\
   & $10.94 \pm 0.07$ & &\\
20 & $10.69 \pm 0.11$ & $116.92 \pm 0.13$ & $127.57 \pm 0.09$\\
\hline
\end{tabular}
\end{center}
\label{table_frequencies}
\end{table}

To confirm that the detrending using a median filter did not suppress any hidden variability within the light curve, we performed a polynomial detrend for each light curve portion. A polynomial fitted the short sub-samples much better it did than the whole light curve. Comparing the results with the previous ones, no significant differences were noted.

\begin{figure}
\begin{center}
\includegraphics[width=90mm]{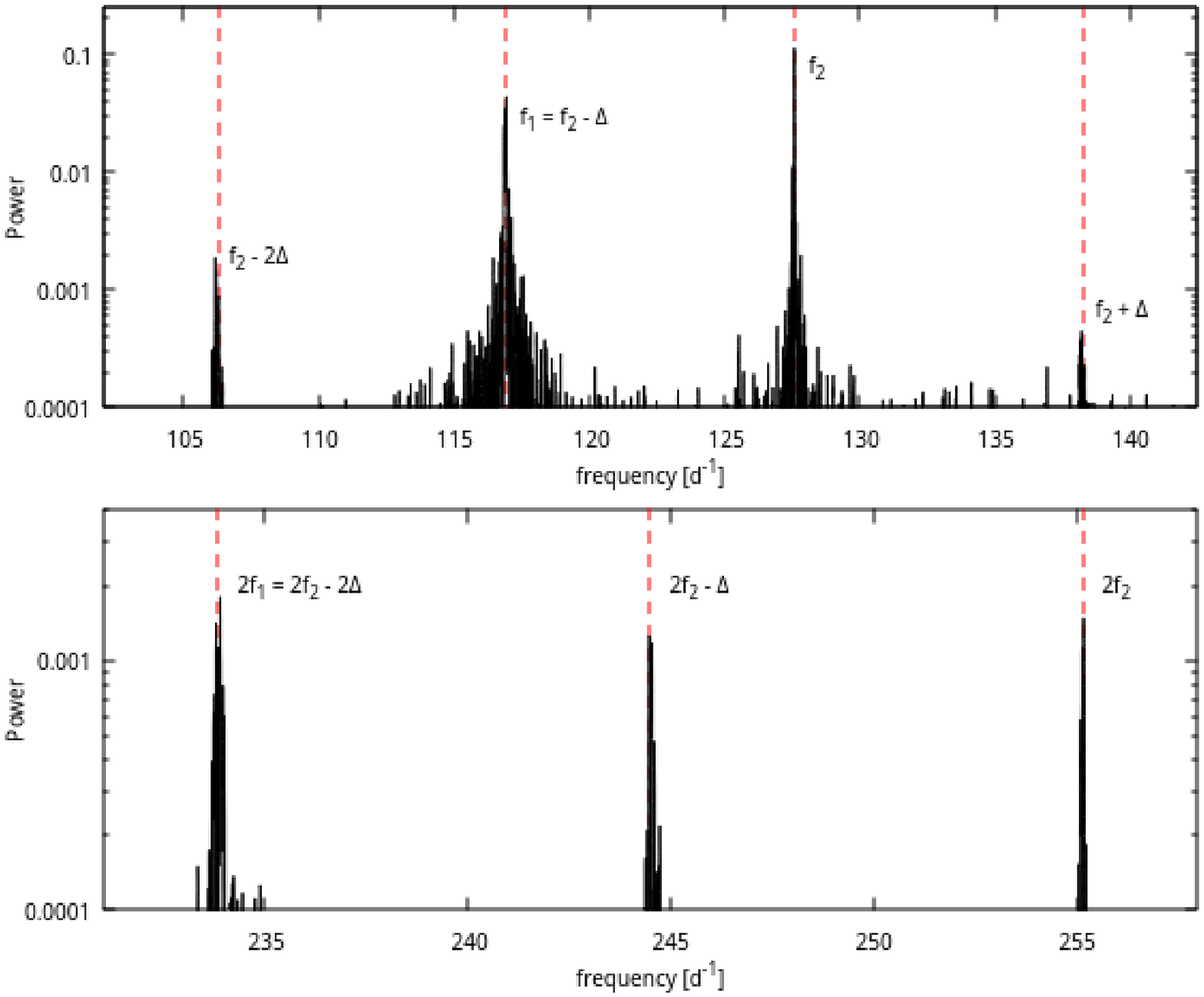}
\caption{LS periodograms. The main frequency $f_2$ and its sidebands are shown on top, and the higher harmonic frequency $2 f_2$ and its sidebands are shown on the bottom panel.}
\label{harmonics_frequencies}
\end{center}
\end{figure}

By inspecting the higher frequencies around 250\,d$^{-1}$ we realize that they may represent higher harmonics. To evaluate this premise, we calculated the higher harmonics as; $2 f_1\approx 234$\,d$^{-1}$ and $2 f_2 \approx 255$\,d$^{-1}$. We subtracted the $\Delta$ value from $2 f_2$ (or added it to $2 f_1$) and obtained a value of 244.5\,d$^{-1}$. By plotting the frequencies together with the calculated values (as the red vertical lines, see Figure \ref{harmonics_frequencies}), we observe that these values match the peaks, supporting our assumption.
\begin{figure}
\begin{center}
\includegraphics[width=90mm]{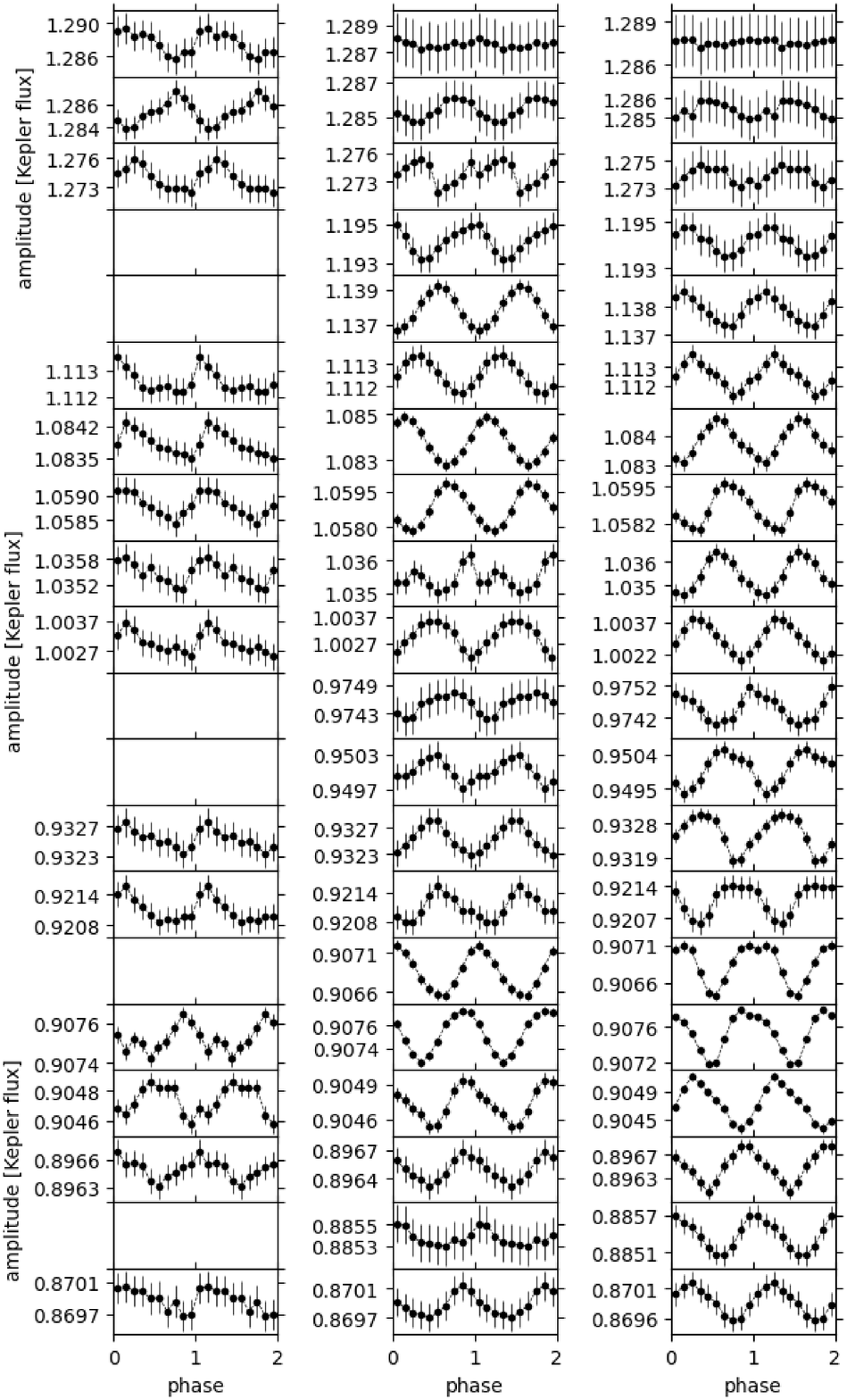}
\caption{Folded light curves of each portion. The first column is folded with frequency $f_0$, the second with $f_1$, and the third with $f_2$. The empty plots for $f_0$ refer to portions with double peak, or to peak with confidence level $<90$\% (see Table \ref{table_frequencies}).}
\label{folded_light_curve}
\end{center}
\end{figure}

Fig.~\ref{folded_light_curve} shows changes in modulation during the observation. The empty blocks for $f_0$ correspond to light curve portions in which the double peak in the LS periodogram was present and an exact measurement was not possible, or the confidence of the peak is below 90\% (see Table \ref{table_frequencies}).

\section{Discussion}
\label{sec_discussion}

The mini-outbursts occurred semi-regularly during the orbital period, for more than 16 years. These flares were peculiar in that they occur once per orbital period, at about the same orbital phase, and kept on recurring. This phenomenon is quite different from other bursting activity of larger amplitude observed in symbiotics \citep[e.g.][]{gromadzki2013}, which does not occur with such clear periodicity. We note that only AG\,Dra has flares with characteristics that are somewhat similar to FN\,Sgr (\citealt{Galis2019}), and in that case they recur with a time scale that is close to the pulsation of the red giant. However, in FN Sgr we do not have high-quality radial velocity data to detect such a pulsation.

The timescale and amplitude are similar to the first flaring event observed in the symbiotic system Z\,And, with 1.5\,mag amplitude, a sharp rise and a decay over $\approx$300 days \citep{sokoloski1999}. However, in Z\,And and other systems with repeated flares (e.g. AX\,Per, CI\,Cyg, BF\,Cyg) the recurrence times were always shorter than the orbital period, usually by about 10-20\% \citep[see e.g. discussion in][]{Mikolajewska1996, Mikolajewska2002}. A new outburst in Z\,And had a larger amplitude (2\,mag) and lasted longer, and actually, there were three separate flare episodes over almost 3 years. \citet{sokoloski2006} attributed this phenomenon outburst to a ``combination nova'', namely nuclear shell burning triggered by a disk instability \citep{sokoloski2006}. An interesting fact is a secondary period of $\simeq$355 days found in the radial velocity data of the giant in Z\,And, attributed to the rotation of the giant \citep{Galis1999, Friedjung2003}, was close to the recurrence semi-period of the outburst.
A combination nova has also been invoked to explain the unusual recent outburst of a CV-like system, V1047\,Cen \citep{Aydi2022}. We also note that in Z\,And a stable oscillation with a 28 min period was detected before and during the flare, which \citet{sokoloski1999} attributed to the rotation of a magnetic WD (with B$\geq 10^5$ Gauss).

The high cadence Kepler light curve of FN\,Sgr over 81 days at the end of 2015 offers new clues and new ``puzzles''. We found a complex periodogram with a low frequency $f_0$, and a group of higher frequencies ($f_1$, $f_2$ and other sidebands), resembling the short orbital periods WD binaries classified as intermediate polars (IPs), where the lower frequency is the orbital frequency of the binary, and the higher frequencies are mainly due to the WD spin and the beating between the spin and the orbital frequency (see e.g. \citealt{ferrario1999}). Other sidebands with higher harmonics can also be present in IPs. By analogy, we suggest that the $f_2$ frequency, which is stable, is likely to be due to the WD rotation. The WD spin frequency in IPs is observable because the polar caps are heated by magnetic accretion funneled to the poles by the magnetic field. Even if an accretion disk is formed, it is disrupted at the magnetospheric radius.

\citet{sokoloski1999} analysed the high-resolution photometry taken in 1998 of the FN\,Sgr and detected a possible variability in the frequency range of $25.5-900$\,d$^{-1}$. Even if the uncertainty is very large, our frequency $f_1$ agrees with this detection indicating the spin rotation was observable also in 1998.

While in IPs there is often a third frequency, corresponding to the beat between the orbital and the rotational period, for FN\,Sgr $f_1$ frequency is the beat between the proposed rotation frequency and the $f_0$ frequency, which is generated by a structure that is not synchronized with the orbital motion. It is likely to be around the WD and to be illuminated by the WD as it rotates. Both $f_0$ and $f_1$ are unstable frequencies, and this is consistent with $f_1$ being the beat because if $f_0$ is unstable, the beating must be unstable too.

The non-stability or quasi-periodicity of $f_0$ is a crucial characteristic to study a physical model. In the periodograms of IPs either the WD spin frequency is dominant, or the beating frequency (\citealt{ferrario1999}). The beat is usually detected when there is an accretion disk, even if it is disrupted at a certain radius, namely in IPs and not in polars. In fact, the WD rotation in polars, which accrete directly via the magnetic stream, tends to be synchronized. We propose two possible explanations for the $f_0$ frequency. Both imply the presence of a ``structure'' or a denser element in the accretion disk. If this has negligible mass compared with the WD, and it orbits a WD of 0.7 M$_\sun$ \citep{Brandi2005} with Keplerian angular velocity derived from the period associated with $f_0$ ($\approx$135.5 minutes), it must be localized in the very inner disk at $\simeq$0.76\,R$_{\rm \odot}$ from the center.

One scenario is that the structure in the disk is not fixed, but is due to rocky detritus around the WD, captured in the accretion disk. This would of course explain why the period is not perfectly stable. Rocky detritus so far has been detected only in 1 out of 3000 WDs, so it is a fairly rare phenomenon \citep[see][]{Vanderburg2020}. We also examined the possibility of a ``dark spot'' like in \citet{Kilic2015}, but that event was recurrent with the period of the WD rotation.

Another possibility is that there is a vertical thickening of the accretion disk, causing variable irradiation and inhomogeneities in the disk itself. In dwarf novae, the observed quasi-periodic oscillations (QPOs) may be caused by vertical thickening of the disc that moves as a travelling wave near the inner edge of the disk, alternately obscuring and reflecting radiation from the disk \citep{woudt2002, warner2002}. Applied to FN\,Sgr, this model means that irradiation by the rotating WD would cause the QPO with beat frequency $f_1$. However, the QPOs in dwarf novae have semi-periods of hundreds of seconds \citep[e.g.][]{woudt2003} and in FN\,Sgr the much longer period would imply inhomogeneities considerably farther from the disk centre compared to dwarf novae. In the model by \citet{warner2002}, winding up and reconnection of magnetic field lines cause inhomogeneities. The magnetic field responsible for the phenomenon may be either that of the WD or that of an equatorial belt on the WD surface (low inertia magnetic accretor model, \citealt{warner2002}). Since the inner disc radius where the inhomogeneities form depends on the magnetic field strength, the WD in FN\,Sgr would have a stronger magnetic field than dwarf novae, consistently with magnetically channeled accretion that allows detection of the rotation period.

\subsection{The possibility of rocky bodies around the WD}

Recent \kepler\ photometry has revealed that WDs have periodic optical variations very similar to the detected 2.2\,h periodicity found in FN\,Sgr. \citet{maoz2015} analysed \kepler\ light curves of 14 hot WDs and detected periodic variations with periodicities from 2 hours to 10 days. Possible explanations include transits of objects of dimensions of the order of $\sim 50$-200\,km. The periodicity may arise from UV metal-line opacity, due to the accretion of rocky material as debris of former planetary systems, a phenomenon observed in many WDs \citep[e.g.][]{jura2003,zuckerman2003,zuckerman2010,vanderburg2015, xu2018,vanderbosch2021}. WD\,2359-434, for instance, shows variability with a period of 2.7\,h \citep{gary2013}.

Some characteristics so far observed for rocky detritus differ significantly from what we detected in FN\,Sgr:
\begin{enumerate}
\item The quasi-periodic signals change in shape and amplitude quite significantly over the course of tens of days, while in FN Sgr the changes in amplitude,
and phase observed over 81 days were much smaller;

\item The shape of the dip is very sharp and a-symmetric. We also note that in WD 1145+017 b, for instance, the small exoplanet orbiting the WD causes a very sharp feature in the lightcurve \citep{vanderburg2015,Gansicke2016}.

\item There are multiple periodic signals with similar periods.

\item Typical timescales of the observed features in the lightcurves are around 1 minute only.
\end{enumerate}

Thus, the dips in luminosity caused by the detritus seem to be always sharp, short in duration, and variable; and often there are multiple periods. Looking at these characteristics, we conclude that our observations do not match the rocky detritus scenario, that thus remains a relatively remote possibility.

\subsection{Interpretation as inhomogeneity in the accretion disk}

Disk inhomogeneities can be generated by the stream disk overflow, causing vertical disk thickening like in GU\,Mus (\citealt{peris2015}). These inhomogeneities may liberate blobs of matter rotating with Keplerian angular velocity at the radius where the vertical thickening occurs. However, in GU\,Mus the stream overflow generates the thickening at the circularisation radius\footnote{The circularisation radius is the distance from the centre where angular momentum from the L$_1$ point equals the local specific angular momentum of a Keplerian disc.} which in FN\,Sgr is approximately 35.2\,R$_{\rm \odot}$, while we inferred a distance of only 0.76\,R$_{\rm \odot}$ from the center.

\citet{lubow1975} estimated the minimum distance of the stream from the centre to be
\begin{equation}
r_{\rm min} = 0.0488\,a\,q^{-0.464},
\end{equation}
where $a$ is the binary separation and $q$ is the binary mass ratio (donor mass divided by the WD). Thus, the minimum distance is 11.8\,R$_{\rm \odot}$ in FN\,Sgr.

These estimates are based on the assumption of a Keplerian disc, but \citet{Brandi2005} suggested that the disk around the WD is geometrically thick. This may mean that in FN Sgr there is a sub-Keplerian advection dominated disk, meaning that the radial velocity of the matter is much larger than in a thin disk (\citealt{narayan1994}). The frequency $f_0$ in thick disks is defined as $\eta/(2\pi)\,\Omega_{\rm K}$, where $\eta$ is the sub-Keplerian factor between approximately 0.2 and 1 (middle panel of Fig.~1 in \citealt{narayan1994}). Thus, the structure or inhomogeneity does not need to be located as far from the WD as in the Keplerian, thin disk case.

While the spin and beat modulations have almost sinusoidal shapes, the QPOs with frequency $f_0$ have a steep rise and slow dissipation (Fig.~\ref{folded_light_curve}) Such asymmetry can be understood as rapid generation of the inhomogeneity that slowly dissipates while orbiting the WD.

We note that, while $f_2$ is clearly detected in almost all sub-samples, the beating frequency $f_1$ is strong only in the first half of the light curve (Fig.~\ref{frequencies_zoom}). If the beat is caused by a body rotating around the WD, the weak power implies that this body is not well irradiated by the rotating WD in the second half of the light curve, or the irradiated surface is not well visible anymore, supporting the idea that this body forms and dissipates slowly during the rotation around the WD.

\begin{figure}
\begin{center}
\includegraphics[width=90mm]{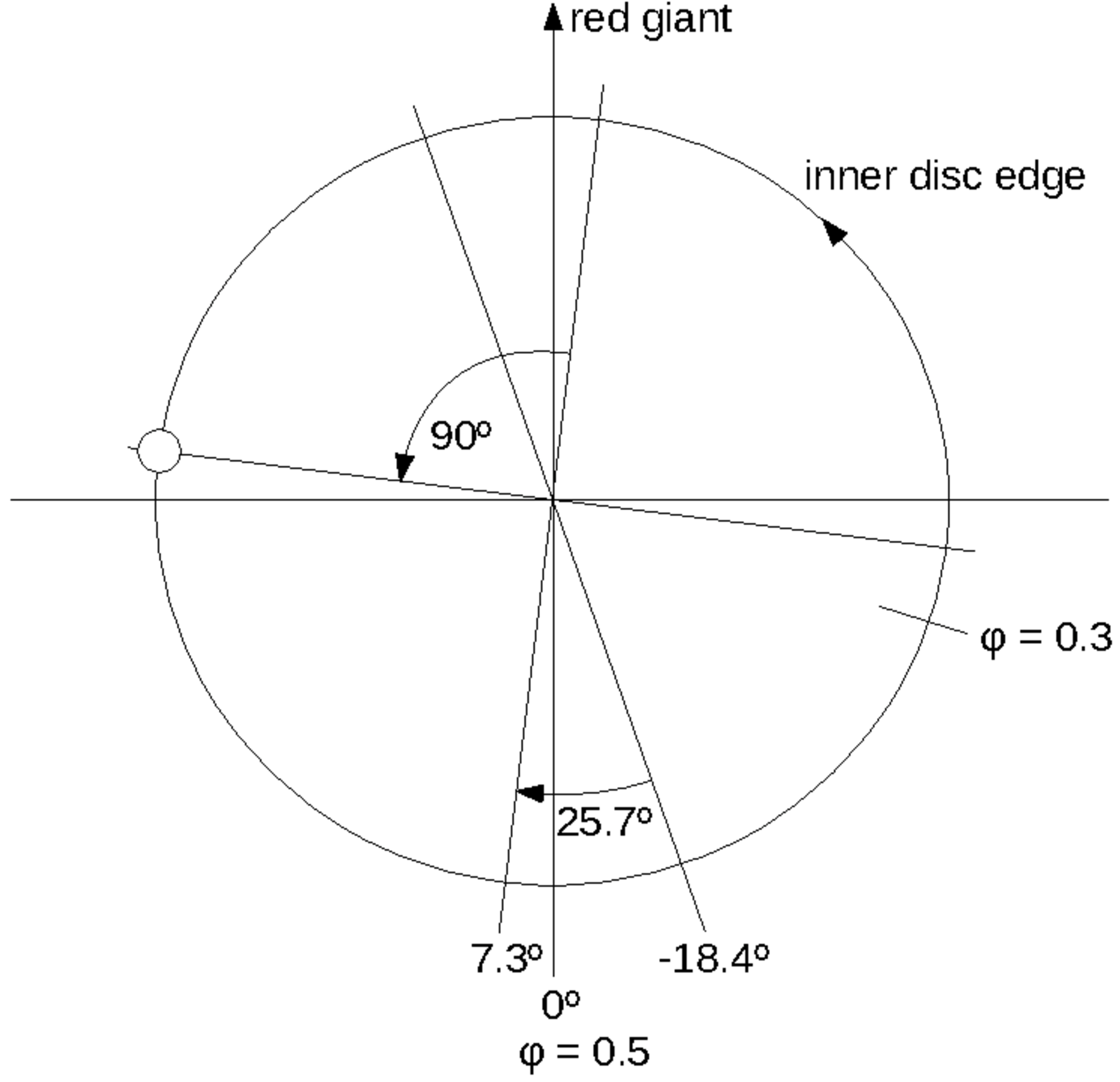}
\caption{Schematic illustration of the angle of view during the \kepler\ observation. The line marked with angle $0^\circ$ points towards the giant or L$_1$. The other angles represent the beginning ($-18.4^\circ$) and the middle ($7.3^\circ$) in the \kepler\ observation. The inner disk edge is marked as a circle with an arrow showing the sense of rotation. The small circle represents the region where the inhomogeneity blobs may be originating (see text for details). $\phi$ denotes the orbital phase and corresponding viewing angle.}
\label{model}
\end{center}
\end{figure}

Since the duration of the light curve is 81 days, the angle of view changed by 0.1428 of the orbital cycle (51.4 degrees). The clear visibility of the $f_1$ frequency during the first half of the light curve suggests that a change in 0.0714 of the orbital cycle (25.7 degrees) modifies the angle of view in such a way that the beating region does not show the irradiated part clearly anymore. Our \kepler\ observation started approximately 29 days before the superior conjunction of the red giant (Fig.~\ref{optical_plus_kepler_2}), which is 0.0511 of the orbital cycle (approximately 18.4 degrees). After half of the observation (25.7 degrees), $f_1$ becomes weaker. Fig.~\ref{model} shows a model of this configuration. The line with angle $-18.4^\circ$ represents the viewing angle at the start of the \kepler\ observation, while $7.3^\circ$ is half of the observation where $f_1$ becomes significantly weaker.

Since at the viewing angle of $7.3^\circ$ the $f_1$ variability starts to decline, if we suppose that there are inhomogeneous blobs visible up to $90^\circ$ from the viewing direction, the small circle in Fig.~\ref{model} shows the regions where these blobs are generated. After sweeping this region, the blobs are not visible enough or dissipate slowly and become too small to generate a strong beat. Following \citet{lubow1975,lubow1976} the region of blobs generation can be associated with the stream disk overflow or simply with the trajectory from the L$_1$ point. Stream disk overflow requires thin disk geometry, while if the disk is geometrically thick, there is not an actual overflow of the accretion stream, but from L$_1$ point the stream may impinge the disk, penetrate it and generate blobs at its inner edge.

%An alternative interpretation of the decreasing power of $f_1$ in the second half of the light curve is that it is not related to the orbital phase, but to the mini-outburst, since we observed it during the decay from the flare peak. We do not have high cadence data outside the flare to test this hypothesis.

\subsection{The mini-outbursts}

Since the outbursts seem to be almost phase-locked, episodic mass accretion rate would be a plausible interpretation. An eccentric orbit could cause mass accretion events when the secondary approaches the primary WD. However, \citet{Brandi2005} concluded that the orbit is almost circular.

An alternative based on a phased locked behaviour is that a bright region appears at specific angles of view. Fig.~\ref{model} shows the angles view of phases 0.3 and 0.5. The mini-outburst ends approximately at phase 0.5, which is very close to the viewing angle where the $f_1$ frequency starts to disappear. This implies a related
physical origin of the two events, and the mini-outburst should be connected to the stream disc overflow or to stream-disc impact region.

Even if the stream-disc scenario seems to be a plausible explanation, we investigated several other possibilities. An explanation in terms of a thermonuclear runaway (a recurrent ``non-ejecting nova'' as in \citet{Yaron2005} implies an accretion rate of order 10$^{-6}$\,M$_\odot$ yr$^{-1}$ and a recurrence time that would decrease from $\simeq$10 years to $\simeq$2 years for increasing WD mass from 0.65 to 1\,M$_\odot$ \citep{Yaron2005}. However, such an outburst would cause an increase by $\simeq$4 magnitudes in V.

A third possibility is that the outbursts are related to the magnetic field. In the old nova and IP GK\,Per, with an orbital period of almost 2 days and a subgiant K2 secondary, dwarf-nova-like-outbursts with a recurrence time of 400$\pm$40 days and amplitude of 1-3\,mag have been observed for decades \citep[see][and references therein]{Zemko2017}. Although the exact mechanism powering these outbursts is a matter of debate, it seems certain that the maximum temperature of the disk and therefore its inner radius decrease in outburst, and more matter is suddenly accreted \citep{Zemko2017}. The time scale of the FN\,Sgr mini-flares is similar, although the much higher luminosity of the system and the larger orbit make it difficult to draw a comparison.

Assuming that the WD is strongly magnetized (B$\geq$10$^5$ Gauss), as our Kepler timing analysis indicates, there are two other, alternative explanations. The magnetosphere may cause ``magnetically gated accretion'' (\citealt{scaringi2017}): unstable, magnetically regulated accretion causing quasi-periodic bursts. In this model, the disc material builds up around the magnetospheric boundary, and reaching a critical amount the matter accretes onto the WD causing an optical flare. \citet{scaringi2017} studied this phenomenon in the cataclysmic variable MV\,Lyr, where quasi-periodic bursts of $\sim 30$\,min appeared every $\sim 2$\,hours. The main question is how the timescale would vary in a symbiotic, a binary with a so much longer orbital period.

The recurrence time of these bursts is typically close to the viscous time-scale $t_{\rm visc}$ in the region where the instability occurs (inner disc);
\begin{equation}
t_{\rm visc} = \frac{r_{\rm in}^2}{\nu},
\end{equation}
where $r_{\rm in}$ is the inner disc radius and $\nu$ is a viscosity parameter. Assuming that the magnetospheric radius 0.76\,R$_{\rm \odot}$ (derived in the previous section), and with the viscosity parameter expressed in term of dimensionless$\alpha$ parameter following \citet{shakura1973} as
\begin{equation}
t_{\rm visc} = \alpha\,(h/r)^2\,(Gm_{\rm WD}r_{\rm in})^{1/2}.
\end{equation}
where $h/r$ is the ratio of the scale height $h$ of the disc at radial distance $r$, $G$ is the gravitational constant and $m_{\rm WD}$ is the WD mass. The mini-outbursts seem to be almost phase locked, therefore for $t_{\rm visc}$ would have to be close to the orbital period of 567.3\, days. If the disc is geometrically thick, with $h/r= 0.1$, the observed recurrence time of FN\,Sgr is obtained only with $\alpha = 0.0026$. However, we know that a realistic value of 0.1 for $\alpha$ in advective disks (\citealt{narayan1994}) yields a much larger magnetospheric radius, namely 8.7\,R$_{\rm \odot}$. Moreover, the $h/r$ ratio may even be as high as 0.4 (e.g. \citealt{godon1996}), shortening $t_{\rm visc}$ considerably.
% It these results were obtained with much higher mass transfer rates that attributed to FN Sgr.
%
On the basis of these considerations, we rule this model out.

An interesting possibility that we would like to consider is that of a localized thermonuclear runaway (LTNR), namely a thermonuclear runaway that does not spread all over the surface, like helium burning on neutron stars. \citet{Shara1982} proposed that dwarf-nova-like, ``vulcanic'' eruptions of small amplitude occur on the surface of massive WDs and may recur on time scales of months, due to LTNRs. In Shara's model, the eruption causes a sort of volcano to reach the WDs surface, without mass ejection. \citep{Orio1993} modelled the LTNR with the formalism used for helium burning on neutron stars and calculated that temperature differences as small as 10$^{4}$-10$^{5}$\,K at the bottom of the envelope accreted by a WD can lead to a LTNR that remains confined and is extinguished before spreading to the all the surface. If accretion funneled by the magnetic field to the WD poles produces such a temperature gradient, a LTNR may occur. The higher the WD mass and the larger the mass accretion rate, the more likely a LTNR would occur. The LTNR may remain confined only for a certain time if it is not extinguished it would later spread to the whole surface, causing a classical nova with a slow rise (a day or more).

More recently, the LTNR scenario has been revisited. A LTNR model involving the magnetic field has been studied for highly magnetized WDs, $B > 10^6$\,G, to explain very small amplitude flashes recently observed in CVs, called micronovae by the authors (\citealt{scaringi2022a, scaringi2022b}). Such flashes cause luminosity increases of a few to $\simeq$30 times occurring with hours and recurring on timescales of months. The strength of the magnetic field that allows accretion to be confined to a region at the poles is estimated assuming the radius of accretion disk truncation by the magnetosphere, caused by magnetic pressure balancing the ram pressure of the accretion flow. This defines a magnetospheric radius (see e.g. \citealt{frank1992})
\begin{equation}
\begin{split}
r_{\rm m} = & 9.8 \times 10^8 \left( \frac{\dot{m}_{\rm acc}}{10^{15}\,{\rm g}\,{\rm s}^{-1}} \right)^{-2/7} \left( \frac{m_{\rm 1}}{{\rm M}_{\rm \odot}} \right)^{-1/7}\\
&\times \left( \frac{\mu}{10^{30}\,{\rm G}\,{\rm cm}^3} \right)^{4/7}\,{\rm cm},
\end{split}
\label{equation_rm}
\end{equation}
where $\dot{m}_{\rm acc}$ is the mass accretion rate, $m_{\rm 1}$ is the WD mass and $\mu = B\,R_{\rm WD}^3$ is the magnetic moment of the WD.

\begin{figure}
\begin{center}
\resizebox{\hsize}{!}{\includegraphics[angle=-90]{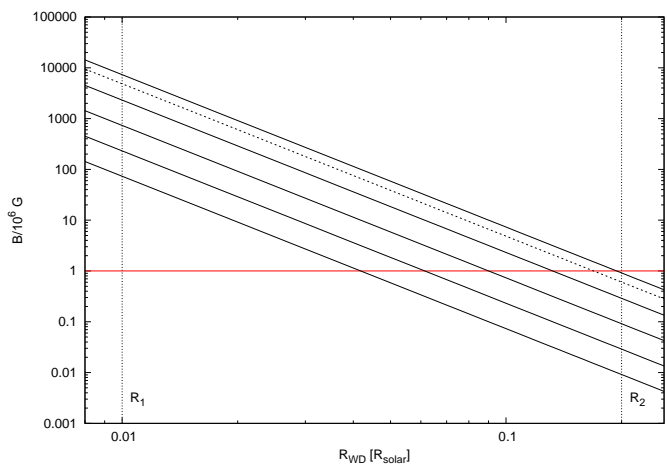}}
\caption{WD magnetic field calculated using Equation~(\ref{equation_rm}) for different values of the mass accretion rate. The solid lines trace the solution with the magnetospheric radius of 0.76\,R$_{\rm \odot}$, and mass accretion rates from 10$^{-6}$ to 10$^{-10}$\,M$_{\rm \odot}$/yr (marked as labels). The dashed line represents the solution with magnetospheric radius 0.6\,R$_{\rm \odot}$ and mass accretion rate 10$^{-6}$\,M$_{\rm \odot}$/yr. The horizontal red line is traced at magnetic field B = 10$^6$\,G. The vertical dotted lines represent two WD radii; $R_1 = 0.013$\,R$_{\rm \odot}$ and $R_2 = 0.02$\,R$_{\rm \odot}$ (see text for details).}
\label{magn_field}
\end{center}
\end{figure}

Fig.~\ref{magn_field} shows $B$ values calculated for various mass accretion rates. As truncation radius $r_{\rm m}$ we assumed 0.76\,R$_{\rm \odot}$. The likely radius of the 0.7 M$_\odot$ WD may be $R_1 = 0.013 R_{\rm \odot}$ (from the mass-radius relation by \citealt{nauenberg1972}). The WD radius may be larger if the WD atmosphere is inflated, but this usually occurs because of hydrogen burning all over the surface \citep[see][]{Starrfield2012}. We take into account the possibility of a larger WD radius in Fig.~\ref{magn_field}. However, we note two issues: a) an apparent typo in \citet{Brandi2005}, since from the values in their Table~5, namely a hot, hydrogen burning WD with T$_{\rm eff}$150,000-180,000 K and L $\simeq$ 1000-2000\,L$_\odot$ should be around R $\simeq$ 0.02\,R$_{\rm \odot}$ instead of 0.2\,R$_{\rm \odot}$, and b) if nuclear burning is localized, the large UV luminosity should be ascribed to the accretion disk and not to the WD. In any case, we take into account the possibility of a larger WD radius in Fig.~\ref{magn_field}.

Allowing for a radius R $\simeq$ 0.02\,R$_{\rm \odot}$, the micronova scenario is acceptable with a magnetic field higher than a few megaGauss with $\dot m$ of a few 10$^{-10}$ M$_\odot$/yr. However, for values of $\dot m$ that are closer to what has been inferred in many symbiotics, the magnetic field should be of at least the order of hundreds of megaGauss, which is unusual for WDs. If the matter orbits more slowly than with the Keplerian velocity because of the geometrically thick disk, the inner disk radius can be smaller, but this would affect the result only slightly, as shown by the dashed line in the Figure.

A detection in supersoft X-rays would be extremely interesting to understand whether the WD atmospheric temperature and bolometric luminosity are as high as derived by \citet{Brandi2005}, in which case the burning is unlikely to be localized. Very constraining upper limits on the bolometric luminosity, however, may be difficult to obtain given the distance and the possible large intrinsic absorption in the symbiotic nebula.

\section{Summary and conclusions}

The V and I light curves observed over decades revealed that for many years, from before 2001 until 2019, the symbiotic underwent recurrent optical flares with amplitude 0.5-1\,mag, at orbital phase 0.3-0.5, with a sharp rise over 10 days and a decay during many weeks. The amplitude and timescales of the outbursts cannot be explained with a disk instability triggered by a burst of mass transfer, neither with a non-ejecting thermonuclear flash nor with a ``combination nova''. Because of the detection of a likely rotation period in the \kepler\ light curve, we found that the WD is likely to have a strong magnetic field, $\geq 10^5$\,Gauss. We examined two alternatives that have recently been proposed for small amplitude flares of magnetic cataclysmic variables, and how they may be relevant for a symbiotic system with a magnetic WD. A thermonuclear runaway in a ``non-ejecting nova'' would occult with higher luminosity amplitude than observed in FN\,Sgr. We found that localized thermonuclear burning confined by the magnetic field may explain the observed phenomena with B $\geq 10^6$ Gauss and accretion rate $\geq 10^{-10}$\,M$_\odot$\,yr$^{-1}$. However, localized burning is difficult to reconcile with the high bolometric luminosity and temperature of the ionizing source, inferred from the UV range and from the flux of the He II $\lambda$ 4686 emission line by \citet{Brandi2005}. If the WD is indeed the ionizing source, as suggested by the above authors, it is so hot and luminous that thermonuclear burning is ongoing all over the surface.

An interesting comparison is the one with the periodic outbursts of the IP and old nova GK\,Per, although the difference in luminosity and spatial dimensions make a rigorous comparison very difficult.

The most promising interpretation is based on the phase locked appearance of the mini-outbursts. It points to the visibility of the stream disc overflow which coincides also with conclusions derived from our detailed timing analysis.

The \kepler\ timing analysis revealed three dominant frequencies with sidebands. The lowest frequency $f_0$ is unstable and probably represents an inhomogeneity generated at the inner disk edge. The higher frequency $f_2$ (11.3\,min periodicity) is very stable and we attribute it to a magnetic rotating WD. We also measured a frequency $f_1$, corresponding to the beating between $f_1$ and $f_2$.

The $f_0$ and $f_2$ frequencies are present during the whole light curve, suggesting permanent visibility of the corresponding sources. The stability of $f_2$ is consistent with the idea that it is the frequency generated by the rotating WD. We interpret $f_0$ as a frequency generated by an inner disk inhomogeneity, indicating the presence of blobs or rigid bodies during the whole orbital period. The light curve folded with the period corresponding to $f_0$ implies that these may be blobs that form and dissipate slowly. The beat frequency $f_1$ is strong only during the first half of the observation. This is consistent with the region of strongest irradiation of the blobs being associated with the trajectory of an accretion stream from the L$_1$ point, impinging the geometrically thick disk of FN\,Sgr, penetrating it and generating inhomogeneity blobs at the inner disk edge.

We examined an alternative explanation for the $f_1$ frequency, namely rocky detritus around the WD, but the characteristics of the modulation are very different from what has been so far observed in nearby WDs.

\section*{Acknowledgement}

JMag, AD, and PB were supported by the European Regional Development Fund, project No. ITMS2014+: 313011W085. JMik was supported by the Polish National Science Centre (NCN) grant OPUS 2017/27/B/ST9/01940.

\bibliographystyle{aa}
\bibliography{mybib}

\begin{thebibliography}{47}
\expandafter\ifx\csname natexlab\endcsname\relax\def\natexlab#1{#1}\fi

\bibitem[{{Aydi} {et~al.}(2022){Aydi}, {Sokolovsky}, {Bright}, {Tremou},
  {Nyamai}, {Evans}, {Strader}, {Chomiuk}, {Myers}, {Hambsch}, {Page},
  {Buckley}, {Woodward}, {Walter}, {Mr{\'o}z}, {Vallely}, {Geballe},
  {Banerjee}, {Gehrz}, {Fender}, {Gromadzki}, {Kawash}, {Knigge}, {Mukai},
  {Munari}, {Orio}, {Ribeiro}, {Sokoloski}, {Starrfield}, {Udalski}, \&
  {Woudt}}]{Aydi2022}
{Aydi}, E., {Sokolovsky}, K.~V., {Bright}, J.~S., {et~al.} 2022, \apj, 939, 6

\bibitem[{{Barba} {et~al.}(1992){Barba}, {Brandi}, {Garcia}, \&
  {Ferrer}}]{barba1992}
{Barba}, R., {Brandi}, E., {Garcia}, L., \& {Ferrer}, O. 1992, \pasp, 104, 330

\bibitem[{{Brandi} {et~al.}(2005){Brandi}, {Miko{\l}ajewska}, {Quiroga},
  {Belczy{\'n}ski}, {Ferrer}, {Garc{\'\i}a}, \& {Pereira}}]{Brandi2005}
{Brandi}, E., {Miko{\l}ajewska}, J., {Quiroga}, C., {et~al.} 2005, \aap, 440,
  239

\bibitem[{{Ferrario} \& {Wickramasinghe}(1999)}]{ferrario1999}
{Ferrario}, L. \& {Wickramasinghe}, D.~T. 1999, \mnras, 309, 517

\bibitem[{{Frank} {et~al.}(1992){Frank}, {King}, \& {Raine}}]{frank1992}
{Frank}, J., {King}, A., \& {Raine}, D. 1992, {Accretion power in
  astrophysics.}, Vol.~21

\bibitem[{{Friedjung} {et~al.}(2003){Friedjung}, {G{\'a}lis}, {Hric}, \&
  {Petr{\'\i}k}}]{Friedjung2003}
{Friedjung}, M., {G{\'a}lis}, R., {Hric}, L., \& {Petr{\'\i}k}, K. 2003, \aap,
  400, 595

\bibitem[{{G{\'a}lis} {et~al.}(1999){G{\'a}lis}, {Hric}, {Friedjung}, \&
  {Petr{\'\i}k}}]{Galis1999}
{G{\'a}lis}, R., {Hric}, L., {Friedjung}, M., \& {Petr{\'\i}k}, K. 1999, \aap,
  348, 533

\bibitem[{{G{\'a}lis} {et~al.}(2019){G{\'a}lis}, {Merc}, {Leedj{\"a}rv},
  {Vra{\v{s}}{\v{t}}{\'a}k}, \& {Karpov}}]{Galis2019}
{G{\'a}lis}, R., {Merc}, J., {Leedj{\"a}rv}, L., {Vra{\v{s}}{\v{t}}{\'a}k}, M.,
  \& {Karpov}, S. 2019, Open European Journal on Variable Stars, 197, 15

\bibitem[{{G{\"a}nsicke} {et~al.}(2016){G{\"a}nsicke}, {Aungwerojwit}, {Marsh},
  {Dhillon}, {Sahman}, {Veras}, {Farihi}, {Chote}, {Ashley}, {Arjyotha},
  {Rattanasoon}, {Littlefair}, {Pollacco}, \& {Burleigh}}]{Gansicke2016}
{G{\"a}nsicke}, B.~T., {Aungwerojwit}, A., {Marsh}, T.~R., {et~al.} 2016,
  \apjl, 818, L7

\bibitem[{{Gary} {et~al.}(2013){Gary}, {Tan}, {Curtis}, {Tristram}, \&
  {Fukui}}]{gary2013}
{Gary}, B.~L., {Tan}, T.~G., {Curtis}, I., {Tristram}, P.~J., \& {Fukui}, A.
  2013, Society for Astronomical Sciences Annual Symposium, 32, 71

\bibitem[{{Godon}(1996)}]{godon1996}
{Godon}, P. 1996, \apj, 462, 456

\bibitem[{{Gromadzki} {et~al.}(2013){Gromadzki}, {Miko{\l}ajewska}, \&
  {Soszy{\'n}ski}}]{gromadzki2013}
{Gromadzki}, M., {Miko{\l}ajewska}, J., \& {Soszy{\'n}ski}, I. 2013, \actaa,
  63, 405

\bibitem[{{Jura}(2003)}]{jura2003}
{Jura}, M. 2003, \apjl, 584, L91

\bibitem[{{Kilic} {et~al.}(2015){Kilic}, {Gianninas}, {Bell}, {Curd}, {Brown},
  {Hermes}, {Dufour}, {Wisniewski}, {Winget}, \& {Winget}}]{Kilic2015}
{Kilic}, M., {Gianninas}, A., {Bell}, K.~J., {et~al.} 2015, \apjl, 814, L31

\bibitem[{{Lubow} \& {Shu}(1975)}]{lubow1975}
{Lubow}, S.~H. \& {Shu}, F.~H. 1975, \apj, 198, 383

\bibitem[{{Lubow} \& {Shu}(1976)}]{lubow1976}
{Lubow}, S.~H. \& {Shu}, F.~H. 1976, \apjl, 207, L53

\bibitem[{{Maoz} {et~al.}(2015){Maoz}, {Mazeh}, \& {McQuillan}}]{maoz2015}
{Maoz}, D., {Mazeh}, T., \& {McQuillan}, A. 2015, \mnras, 447, 1749

\bibitem[{{Mikolajewska}(1996)}]{Mikolajewska1996}
{Mikolajewska}, J. 1996, in Astrophysics and Space Science Library, Vol. 208,
  IAU Colloq. 158: Cataclysmic Variables and Related Objects, ed. A.~{Evans} \&
  J.~H. {Wood}, 335

\bibitem[{Mikolajewska(2002)}]{Mikolajewska2002}
Mikolajewska, J. 2002

\bibitem[{{Miko{\l}ajewska}(2012)}]{mikolajewska2012}
{Miko{\l}ajewska}, J. 2012, Baltic Astronomy, 21, 5

\bibitem[{{Munari} \& {Buson}(1994)}]{munari1994}
{Munari}, U. \& {Buson}, L.~M. 1994, \aap, 287, 87

\bibitem[{{Narayan} \& {Yi}(1994)}]{narayan1994}
{Narayan}, R. \& {Yi}, I. 1994, \apjl, 428, L13

\bibitem[{{Nauenberg}(1972)}]{nauenberg1972}
{Nauenberg}, M. 1972, \apj, 175, 417

\bibitem[{{Orio} \& {Shaviv}(1993)}]{Orio1993}
{Orio}, M. \& {Shaviv}, G. 1993, \apss, 202, 273

\bibitem[{{Peris} {et~al.}(2015){Peris}, {Vrtilek}, {Steiner}, {Vrtilek}, {Wu},
  {McClintock}, {Longa-Pe{\~n}a}, {Steeghs}, {Callanan}, {Ho}, {Orosz}, \&
  {Reynolds}}]{peris2015}
{Peris}, C.~S., {Vrtilek}, S.~D., {Steiner}, J.~F., {et~al.} 2015, \mnras, 449,
  1584

\bibitem[{{Scargle}(1982)}]{scargle1982}
{Scargle}, J.~D. 1982, \apj, 263, 835

\bibitem[{{Scaringi} {et~al.}(2022{\natexlab{a}}){Scaringi}, {Groot}, {Knigge},
  {Bird}, {Breedt}, {Buckley}, {Cavecchi}, {Degenaar}, {de Martino}, {Done},
  {Fratta}, {I{\l}kiewicz}, {Koerding}, {Lasota}, {Littlefield}, {Manara},
  {O'Brien}, {Szkody}, \& {Timmes}}]{scaringi2022a}
{Scaringi}, S., {Groot}, P.~J., {Knigge}, C., {et~al.} 2022{\natexlab{a}},
  \nat, 604, 447

\bibitem[{{Scaringi} {et~al.}(2022{\natexlab{b}}){Scaringi}, {Groot}, {Knigge},
  {Lasota}, {de Martino}, {Cavecchi}, {Buckley}, \&
  {Camisassa}}]{scaringi2022b}
{Scaringi}, S., {Groot}, P.~J., {Knigge}, C., {et~al.} 2022{\natexlab{b}},
  \mnras, 514, L11

\bibitem[{{Scaringi} {et~al.}(2017){Scaringi}, {Maccarone}, {D'Angelo},
  {Knigge}, \& {Groot}}]{scaringi2017}
{Scaringi}, S., {Maccarone}, T.~J., {D'Angelo}, C., {Knigge}, C., \& {Groot},
  P.~J. 2017, \nat, 552, 210

\bibitem[{{Shakura} \& {Sunyaev}(1973)}]{shakura1973}
{Shakura}, N.~I. \& {Sunyaev}, R.~A. 1973, \aap, 24, 337

\bibitem[{{Shara}(1982)}]{Shara1982}
{Shara}, M.~M. 1982, \apj, 261, 649

\bibitem[{{Sokoloski} \& {Bildsten}(1999)}]{sokoloski1999}
{Sokoloski}, J.~L. \& {Bildsten}, L. 1999, \apj, 517, 919

\bibitem[{{Sokoloski} {et~al.}(2006){Sokoloski}, {Kenyon}, {Espey}, {Keyes},
  {McCandliss}, {Kong}, {Aufdenberg}, {Filippenko}, {Li}, {Brocksopp},
  {Kaiser}, {Charles}, {Rupen}, \& {Stone}}]{sokoloski2006}
{Sokoloski}, J.~L., {Kenyon}, S.~J., {Espey}, B.~R., {et~al.} 2006, \apj, 636,
  1002

\bibitem[{{Starrfield} {et~al.}(2012){Starrfield}, {Iliadis}, {Timmes}, {Hix},
  {Arnett}, {Meakin}, \& {Sparks}}]{Starrfield2012}
{Starrfield}, S., {Iliadis}, C., {Timmes}, F.~X., {et~al.} 2012, Bulletin of
  the Astronomical Society of India, 40, 419

\bibitem[{{Vanderbosch} {et~al.}(2021){Vanderbosch}, {Rappaport}, {Guidry},
  {Gary}, {Blouin}, {Kaye}, {Weinberger}, {Melis}, {Klein}, {Zuckerman},
  {Vanderburg}, {Hermes}, {Hegedus}, {Burleigh}, {Sefako}, {Worters}, \&
  {Heintz}}]{vanderbosch2021}
{Vanderbosch}, Z.~P., {Rappaport}, S., {Guidry}, J.~A., {et~al.} 2021, \apj,
  917, 41

\bibitem[{{Vanderburg} \& {Johnson}(2014)}]{vanderburg2014}
{Vanderburg}, A. \& {Johnson}, J.~A. 2014, \pasp, 126, 948

\bibitem[{{Vanderburg} {et~al.}(2015){Vanderburg}, {Johnson}, {Rappaport},
  {Bieryla}, {Irwin}, {Lewis}, {Kipping}, {Brown}, {Dufour}, {Ciardi}, {Angus},
  {Schaefer}, {Latham}, {Charbonneau}, {Beichman}, {Eastman}, {McCrady},
  {Wittenmyer}, \& {Wright}}]{vanderburg2015}
{Vanderburg}, A., {Johnson}, J.~A., {Rappaport}, S., {et~al.} 2015, \nat, 526,
  546

\bibitem[{{Vanderburg} {et~al.}(2016){Vanderburg}, {Latham}, {Buchhave},
  {Bieryla}, {Berlind}, {Calkins}, {Esquerdo}, {Welsh}, \&
  {Johnson}}]{vanderburg2016}
{Vanderburg}, A., {Latham}, D.~W., {Buchhave}, L.~A., {et~al.} 2016, \apjs,
  222, 14

\bibitem[{{Vanderburg} {et~al.}(2020){Vanderburg}, {Rappaport}, {Xu},
  {Crossfield}, {Becker}, {Gary}, {Murgas}, {Blouin}, {Kaye}, {Palle}, {Melis},
  {Morris}, {Kreidberg}, {Gorjian}, {Morley}, {Mann}, {Parviainen}, {Pearce},
  {Newton}, {Carrillo}, {Zuckerman}, {Nelson}, {Zeimann}, {Brown},
  {Tronsgaard}, {Klein}, {Ricker}, {Vanderspek}, {Latham}, {Seager}, {Winn},
  {Jenkins}, {Adams}, {Benneke}, {Berardo}, {Buchhave}, {Caldwell},
  {Christiansen}, {Collins}, {Col{\'o}n}, {Daylan}, {Doty}, {Doyle},
  {Dragomir}, {Dressing}, {Dufour}, {Fukui}, {Glidden}, {Guerrero}, {Guo},
  {Heng}, {Henriksen}, {Huang}, {Kaltenegger}, {Kane}, {Lewis}, {Lissauer},
  {Morales}, {Narita}, {Pepper}, {Rose}, {Smith}, {Stassun}, \&
  {Yu}}]{Vanderburg2020}
{Vanderburg}, A., {Rappaport}, S.~A., {Xu}, S., {et~al.} 2020, \nat, 585, 363

\bibitem[{{Warner} \& {Woudt}(2002)}]{warner2002}
{Warner}, B. \& {Woudt}, P.~A. 2002, \mnras, 335, 84

\bibitem[{{Woudt} \& {Warner}(2002)}]{woudt2002}
{Woudt}, P.~A. \& {Warner}, B. 2002, \mnras, 333, 411

\bibitem[{{Woudt} \& {Warner}(2003)}]{woudt2003}
{Woudt}, P.~A. \& {Warner}, B. 2003, \mnras, 340, 1011

\bibitem[{{Xu} {et~al.}(2018){Xu}, {Rappaport}, {van Lieshout}, {Vanderburg},
  {Gary}, {Hallakoun}, {Ivanov}, {Wyatt}, {DeVore}, {Bayliss}, {Bento},
  {Bieryla}, {Cameron}, {Cann}, {Croll}, {Collins}, {Dalba}, {Debes}, {Doyle},
  {Dufour}, {Ely}, {Espinoza}, {Joner}, {Jura}, {Kaye}, {McClain}, {Muirhead},
  {Palle}, {Panka}, {Provencal}, {Randall}, {Rodriguez}, {Scarborough},
  {Sefako}, {Shporer}, {Strickland}, {Zhou}, \& {Zuckerman}}]{xu2018}
{Xu}, S., {Rappaport}, S., {van Lieshout}, R., {et~al.} 2018, \mnras, 474, 4795

\bibitem[{{Yaron} {et~al.}(2005){Yaron}, {Prialnik}, {Shara}, \&
  {Kovetz}}]{Yaron2005}
{Yaron}, O., {Prialnik}, D., {Shara}, M.~M., \& {Kovetz}, A. 2005, \apj, 623,
  398

\bibitem[{{Zemko} {et~al.}(2017){Zemko}, {Orio}, {Luna}, {Mukai}, {Evans}, \&
  {Bianchini}}]{Zemko2017}
{Zemko}, P., {Orio}, M., {Luna}, G.~J.~M., {et~al.} 2017, \mnras, 469, 476

\bibitem[{{Zuckerman} {et~al.}(2003){Zuckerman}, {Koester}, {Reid}, \&
  {H{\"u}nsch}}]{zuckerman2003}
{Zuckerman}, B., {Koester}, D., {Reid}, I.~N., \& {H{\"u}nsch}, M. 2003, \apj,
  596, 477

\bibitem[{{Zuckerman} {et~al.}(2010){Zuckerman}, {Melis}, {Klein}, {Koester},
  \& {Jura}}]{zuckerman2010}
{Zuckerman}, B., {Melis}, C., {Klein}, B., {Koester}, D., \& {Jura}, M. 2010,
  \apj, 722, 725

\end{thebibliography}

% \label{lastpage}

\end{document}